\documentclass[aps,pre,twocolumn,epsfig]{revtex4}
\usepackage{graphicx}
\usepackage{amsfonts}
\newcommand{\ds}{\displaystyle}

\begin{document}
\title{Gap solitons in Bose-Einstein condensates in  linear and nonlinear optical lattices}
\author{Fatkhulla  Abdullaev\dag,\ddag\
Abdulaziz Abdumalikov\ddag\ and Ravil Galimzyanov\ddag }
\affiliation{\dag\ Instituto de Fisica Teorica, UNESP, Rua
Pamplona, 145, 01405-900, Sao Paulo, SP, Brasil\\
\ddag\ Physical-Technical Institute of the Academy of Sciences,
700084, Tashkent-84, G.Mavlyanov str.,2-b, Uzbekistan }

\begin{abstract}
Properties of localized states on array of BEC confined to a
potential, representing superposition of linear and nonlinear
optical lattices are investigated. For a shallow lattice case  the
coupled mode system has been derived. The modulational instability
of nonlinear plane waves is analyzed. We revealed new types of gap
solitons and studied their stability. For the first time a moving
soliton solution has been found. Analytical predictions are
confirmed by numerical simulations of the Gross-Pitaevskii
equation with jointly acting linear and nonlinear periodic
potentials.
\end{abstract}
\maketitle

\newpage

\section{Introduction}
The Bose-Einstein condensate in a linear periodic potential
attracts a great attention for last years. Many fascinating
phenomena like Josephson oscillations, macroscopic quantum
tunnelling and localization, gap solitons etc have been predicted
and observed in the experiments \cite{Morsh,BK}. The description
of these  phenomena is based on the Gross-Pitaevskii equation
governing the condensate wave function  and having terms
corresponding to the external linear (varying in space and time)
potential and a mean field nonlinearity(taking into account many
body effects). The strength of the mean field nonlinearity is
proportional to the atomic scattering length. We can imagine the
BEC system, when the strength of two body interaction is varying
in space\cite{AS03,kevrekidis,AGKT,GA}. For a periodic variation
of scattering length it leads to appearance of a nonlinear
periodic potential. The ground state and dynamics of localized
states of BEC under action of the nonlinear periodic potential
have been studied recently in papers \cite{SM,AG,Fibich}.

Periodic potential in BEC\cite{OFR1,OFR2} can be generated by
optical methods.  Two types of optical lattices are considered: a
linear periodic potential induced by the standing laser field
\cite{Morsh} and a nonlinear optical lattice produced by two
counter propagating laser beams with parameters near the optically
induced Feshbach resonance (FR) \cite{SM,AG}. Periodic variation
of the laser field intensity in space
by proper choice of the resonance detuning
can lead to the spatial dependence of the scattering length
\cite{OFR2}. Then the nonlinear
term in the GP equation becomes periodically modulated in space,
leading to the generation of nonlinear optical lattice.

In real physical situations exact resonance condition is not
attained and the potential represents the mixture of linear and
nonlinear optical lattices. This type of periodic potential also
can be produced by two pairs of counter propagating laser  beams,
when one pair produces linear optical lattice, and second pair - a
nonlinear one. Another interesting system where the superposition
of linear and nonlinear periodic potentials can be realized is the
array of fermion-boson mixtures\cite{Konotop06}. Gap solitons(GS)
in a photorefractive crystal, considered recently in
\cite{Malomed05},
also originate from the jointly acting linear and nonlinear
periodic potentials. But the last system is different from ones
considered here, since its nonlinearity has saturable nature.

Two types of optical lattices can be distinguished: a shallow and
a deep. The first type is realized for $V< E_R$, and the second
one for $V
> 5E_R$, where $V$ is the strength of the periodic potential and
$E_R = \hbar^2 k^2/2m$ is the recoil energy, $k$ is the laser beam
wavenumber.

In the case of a shallow optical lattice (that will be considered
in this work) we can apply a coupled mode approach. The
coupled-mode theory describes general properties of gap solitons,
taking into account interaction of counter propagating waves.
Analytical solutions have been found for the case of the Kerr
nonlinearity in \cite{AW}. In our case the system of coupled-mode
theory  has more complicated character and is analogous to the
studied in the nonlinear optics for a deep Bragg grating
case\cite{Iizuka} and a nonlinear layered structures\cite{Pel1}.
In the case of
vanishing nonlinear lattice a periodic potential transforms into
the standard linear one \cite{Sterke,YS,EC}. We study the
modulational instability(MI) of nonlinear plane waves which is
important mechanism for the soliton and solitonic pattern
generation. The gap soliton solutions of this modified coupled
mode system are obtained and their stability is analyzed.

The emphasis is also given to the travelling gap soliton (GS)
solutions. We obtain solutions for moving GS and study collisions
between solitons having an inelastic character. In optics they
have been experimentally observed  in fibers with Bragg
grating\cite{Eggleton}. It will be interesting to observe the
travelling GS in BEC with linear and nonlinear optical lattices.

The case of deep optical lattice can be investigated in the
framework of the tight-binding approximation. The GP equation then
transforms into the modified discrete nonlinear Schr\"odinger
equation (NLSE) with a nonlocal nonlinearity\cite{TS,ABDKS,Khare}.
Analysis of MI and the discrete breathers for this model requires
separate investigation.

The paper is organized as follows. In Section II  we describe the
model and derive the coupled mode system. In Section III
modulational instability of cw solutions is investigated. The gap
soliton solutions and their stability is discussed in Section IV.
The moving gap soliton solutions and the solitons collision are
studied in Section V. Main results of the paper are summarized in
Section VI.

\section{The model. Coupled-mode equations}
We consider here the dynamics of BEC under jointly acting linear
and nonlinear optical lattices. First type of potential is induced
by a periodic interference pattern from the counter propagating
laser beams\cite{Morsh}. Second type of potential is produced by
the optically induced Feshbach resonance \cite{OFR2}. According to
this technique, periodic variation of laser field intensity in the
standing wave $I(x) = I_{0}\cos^{2}(kx)$ produces periodic
variation of the atomic scattering length $a_{s}$, i.e.
\begin{equation}
a_{s} = a_{s0} + \alpha \frac{I(x)}{\delta + \alpha I},
\end{equation}
where $\delta$ is the detuning. At large detunings from FR $a_{s}
= a_{s0} + a_{s1}\cos^{2}(kx) $.

In the recent experiment with the $^{87}$Rb condensate the
scattering length has been optically manipulated\cite{OFR1}. The
number of of atoms was $10^6$ and laser power was approximately
$500 W/cm^2$. By choice of the laser power and detuning of the
laser beam around the photo-association resonance the scattering
length was changed over one order of magnitude, from $10 a_0$ to
$190 a_0$ ($a_0$ is the Bohr radius).

  The quasi 1D GP equation for the BEC wavefunction under action
of superposition of a linear and nonlinear optical lattice is
\cite{SM,AG}:
  \begin{eqnarray}
   i\hbar\psi_T = -\frac{\hbar^2}{2m}\psi_{XX} +
   V_{0}\cos^{2}(kx)\psi + (g_{1D}^{(0)} \nonumber\\ +  g_{1D}^{(1)} \cos^{2}(kx))|\psi|^2 \psi,
  \end{eqnarray}
where $g_{1D}^{(0,1)}=2\hbar a_{s0,1}\omega_{\perp}$, and $a_s$ is
the atomic scattering length. Introducing dimensionless variables
\begin{eqnarray*}
\epsilon = \frac{V_0}{2E_R}, \  E_R = \frac{\hbar^2k^2}{2m}, \ x =
Xk, \ t =
\frac{T}{T_{0}}, \\
T_{0} = \frac{E_{R}}{\hbar}, \ u =
\sqrt{\frac{2|a_{s0}|\omega_{\perp}}{E_{R}}}\psi e^{-i\epsilon t},
\ \kappa = \frac{a_{s1}}{a_{s0}}.
\end{eqnarray*} we
can rewrite the equation in  the form
  \begin{equation} \label{GP}
   iu_t + u_{xx} + (\gamma_{0} + \kappa\cos(2x))|u|^2 u - \epsilon\cos(2x)u  =
   0.
  \end{equation}

$\gamma_0 > 0$ corresponds to the attractive and $\gamma_0 < 0$ -
to the repulsive condensates respectively. We will consider the
shallow lattice case, when $\epsilon \ll 1$. In this case one can
derive the system of coupled mode equations. Let us represent the
field $u(x,t)$ in the form
  \begin{equation}
   u(x,t) = (A(x,t)e^{ix} +  B(x,t)e^{-ix})e^{-it}.
  \end{equation}
Collecting terms at $e^{\pm ix}$ we obtain the system of coupled
mode equations
  \begin{eqnarray}\label{cm}
   iA_t + 2iA_x+\gamma_0(|A|^2 A + 2|B|^2 A)\nonumber\\
     -\frac{\epsilon}{2}B + \frac{\kappa}{2}(|B|^2 B +
     2|A|^2 B+A^2B^{\ast})=0,\\ \label{cm1}
   iB_t - 2iB_x+\gamma_0(|B|^2 B + 2|A|^2 B)\nonumber\\
     -\frac{\epsilon}{2}A + \frac{\kappa}{2}(|A|^2 A +
     2|B|^2 A + B^2 A^{\ast})= 0.
  \end{eqnarray}

  In comparison with the standard coupled-mode system,
  this system includes  terms corresponding to the nonlinear
  coupling $\sim \kappa$.

  Note that at small amplitudes of  waves $A,B \sim \exp(ikx - i\omega t)$
  the dispersion law is
$\omega^2 = \epsilon^2/4 + k^2 $, and this system has the
forbidden gap for $ -\epsilon/2 < \omega < \epsilon/2$. The system
is unchanged if we perform transformations $\gamma_0 \rightarrow
-\gamma_0, \  \kappa \rightarrow \kappa, \  \ (A,B, \omega )
\rightarrow (A^{\ast}, -B^{\ast}, -\omega)$.

It means that  results obtained for the first excited band $\omega
> \epsilon/2$ with repulsive interactions can be valid for the
zero order allowed band $\omega < -\epsilon/2$ with attractive
interactions\cite{YS}.

The Eqs. (\ref{cm}), (\ref{cm1}) read as
$$iA(B)_t = \frac{\delta H}{\delta A^{\ast}(B^{\ast})}.$$ The
Hamiltonian $H$ has the following form
  \begin{eqnarray}\label{ham}
   H=\int dx[\frac{i}{2}(A^{\ast}A_x-AA^{\ast}_x-B^{\ast}B_x+
   BB^{\ast}_x) - \nonumber\\
   \frac{\epsilon}{2}(BA^{\ast} + A B^{\ast}) +
   \frac{\gamma_0}{2}(4|A|^2 |B|^2 + \nonumber\\
   |A|^4 + |B|^4 )+ \frac{\kappa}{2}(|A|^2 +
   |B|^2)( BA^{\ast} +B^{\ast}A)].
  \end{eqnarray}

The number of atoms is conserved
  \begin{equation}\label{norm}
   N = \int_{-\infty}^{\infty}dx [|A|^2 + |B|^2].
  \end{equation}

\section{Modulational instability}

We start with consideration of the modulational instability(MI)
phenomena in this model.
The MI is a general phenomenon originating from the interplay
between nonlinearity and dispersion and it consists in the
instability of nonlinear plane wave solutions against long
wavelength modulations. In periodic media the periodicity leads to
modification of the dispersion law. MI is known to be an important
mechanism for the generation of solitons and solitonic trains in
nonlinear optics and Bose-Einstein condensates
\cite{Sterke2,ADG,KS}. The MI in photonic crystals with a
deep Bragg grating recently has been investigated numerically in
Ref.\cite{Porz}. The cw solutions of the coupled-mode system are
sought using the parametrization suggested in \cite{Sterke2}
  \begin{equation}
    A = \frac{\alpha  }{\sqrt{1+f^2}} e^{i(Qx -\Omega t)}, \
    B = \frac{\alpha f}{\sqrt{1+f^2}} e^{i(Qx -\Omega t)},
  \end{equation}
where $\alpha^2 = |A|^2 + |B|^2$ and we have
  \begin{eqnarray}
   \Omega &=& -\frac{3\gamma_0}{2}\alpha^2 -
    \frac{\epsilon}{4}\frac{1+f^2}{f}- \frac{\kappa
    \alpha^2}{4}\frac{1+6f^2+f^4}{f(1+f^2)},\\
   Q &=& -\frac{\gamma_0 \alpha^2}{4}\frac{1-f^2}{1+f^2} +
    \frac{\epsilon}{8}\frac{1-f^2}{f} -
    \frac{\kappa\alpha^2}{8}\frac{1-f^2}{f}.
  \end{eqnarray}
When $\kappa = 0$ the result for a standard coupled-mode system is
reproduced. The physical mean of the parameter $f$: $f < 0$
corresponds to the top of a band gap of the linear dispersion
curve, and $f > 0$ to the bottom of a band gap .

To study MI we look for the solution of the form
  \begin{eqnarray}
   A=\left(\frac{\alpha}{\sqrt{1+f^2}}+
    \delta A(x,t)\right)e^{i(Qx-\Omega t)},\nonumber\\
   B=\left(\frac{\alpha f}{\sqrt{1+f^2}}+
    \delta B(x,t)\right)e^{i(Qx-\Omega t)},
  \end{eqnarray}
where $\delta A$ and $\delta B$ are perturbations of the cw
solution.
Substituting these expressions into Eqs.(\ref{cm}) and
using a linear approximation,  we get the system for $\delta
A(x,t), \delta B(x,t)$
  \begin{eqnarray} \label{syscorr}
   i\delta A_t + 2i\delta A_x + \frac{\epsilon}{2}f\delta A -
     \frac{\epsilon}{2}\delta B + \nonumber\\
     G[\delta A +\delta A^{\ast} + 2f(\delta B + \delta B^{\ast})]+
     \nonumber\\
     K\left[\delta B^{\ast}+2\delta B+\frac{f(1-f^2)}{1+f^2}\delta A+
     \frac{2f}{1+f^2}\delta A^{\ast}\right]=0, \nonumber\\
   i\delta B_t - 2i\delta B_x + \frac{\epsilon}{2f}\delta B -
     \frac{\epsilon}{2}\delta A +  \nonumber\\
     G[f^2 (\delta B^{\ast}+\delta B)+2f(\delta A+\delta A^{\ast})]+
     \nonumber\\
     K\left[\delta A^{\ast}+2\delta A-\frac{1-f^2}{f(1+f^2)}\delta B+
    \frac{2f}{1+f^2}\delta B^{\ast} \right] =0.
  \end{eqnarray}
Here $G = \gamma_0 \alpha^2 /(1+f^2), \ K = \kappa \alpha^2/2$.
The solution of Eqs. (\ref{syscorr}) can be sought in the form
  \begin{equation}
   \delta A(B)=C(D)\cos(qx-\omega t)+iE(F)\sin(qx-\omega t).
  \end{equation}

Substituting these expressions into the system (\ref{syscorr}), we
get the characteristic determinant for it. The full expression is
cumbersome and only some numerical simulations has been performed
for the photonic crystal case\cite{Porz}.

Here we perform the analytical consideration, for the case $f=\pm
1$.

For Eqs.(\ref{syscorr}) we obtain the following characteristic
equation


 \begin{eqnarray}
   \omega^4-\left[(\epsilon-2K)^2-2fG(\epsilon-2K)+
   8q^2\right]\omega^2-\nonumber\\
   8q^2(2K+3fG)(\epsilon-2K)+16q^4 = 0.
  \end{eqnarray}
It should be reminded that here $|f| = 1$.

%
Let us rewrite this equation in the form
  \begin{eqnarray}\label{charac}
   \omega^4 - (8q^2+N_{0})\omega^2-8M_{0}q^2+16q^4=0
  \end{eqnarray}
where
  $$
   N_{0}=(\epsilon-2K)^2-2fG(\epsilon-2K),
  $$
  $$
   M_{0}=(2K+3fG)(\epsilon-2K).  $$

$f=1$ and $f=-1$ correspond to the bottom  (negative effective
mass) and the top (positive effective mass) of the band gap,
respectively, The coefficients of Eq. (\ref{charac}) are real and
this equation can be solved analytically
  \begin{equation}\label{root1}
   \omega^2_1= \frac{1}{2}\left[N_{0} +8q^2+
   \sqrt{N_{0}^2 + 16(N_{0} +2M_{0})q^2}\right],
  \end{equation}
  \begin{equation}\label{root2}
   \omega^2_2= \frac{1}{2}\left[N_{0} +8q^2-
   \sqrt{N_{0}^2 + 16(N_{0} +2M_{0})q^2}\right].
  \end{equation}

Complex roots of Eqs. (\ref{root1}), (\ref{root2}) correspond to
the MI of cw solutions.
Analysis of these equations shows, existence of complex roots is
possible in two cases:

$ q^2 > -R_0, \quad R_0<0$ \\
and

$ q^2 <  M_0, \quad \ M_0>0$ \\
where $$ R_0=\frac{N_0^2}{16(N_0+2M_0)}.$$
Let us analyze these MI
conditions depending on the nonlinear coupling coefficient,
$\kappa$. There is a threshold value $\kappa_0=\epsilon/\alpha^2$,
when $N_0=0$, \  $M_0=0$, \ $R_0=0$ and the MI condition $ q^2 >
-R_0$ is replaced by $ q^2 > M_0$.

We consider two cases:

a) $f = 1$ The root  $\omega_2$ is complex at
   $$
    \kappa>\kappa_0, \quad q^2>-R_0;\quad
    \kappa<\kappa_0, \quad q^2<M_0.
   $$


b) $f = -1$ The root  $\omega_2$ is complex at
   $$
    \kappa>\kappa_0+\gamma_0, \quad q^2>-R_0;
   $$
   $$
    \kappa_0+\gamma_0<\kappa<\kappa_0, \quad -R_0<q^2<M_0;
   $$
   $$
    \kappa<\kappa_0 \quad q^2>-R_0.
   $$


As to the root $\omega_1$ it is complex for all values of $\kappa$
at $\quad q^2>-R_0$ except two points $\kappa=\kappa_0 + \gamma_0$
and $\kappa = \kappa_0$.



When $f = 1$ and $\kappa =0$ the instability of cw exists in the
interval
  $$
   -\sqrt{\frac{3}{2}\epsilon G}< q < \sqrt{\frac{3}{2}\epsilon G},
  $$
that is the same as the result obtained for the standard coupled
mode system \cite{Sterke2}.

\begin{figure}
\centerline{\includegraphics[width=2.5in,angle=-90]{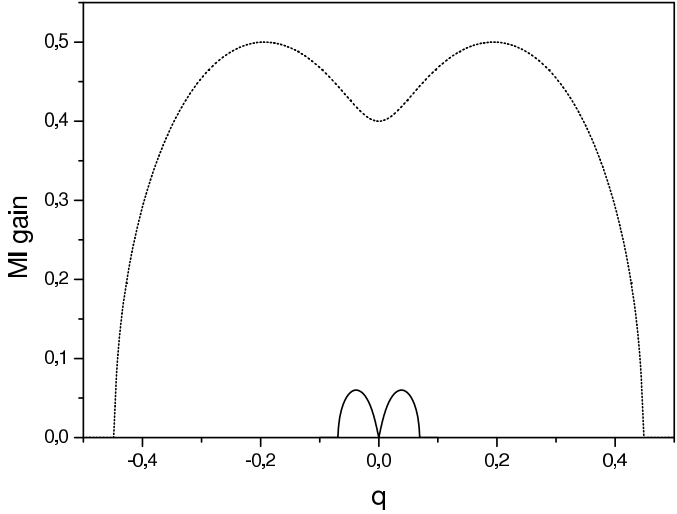}}
\caption{MI gain versus the wavenumber of modulations $q$.
Parameters are $\kappa=1, \ \epsilon = 0.2, \
 f= -1$. Solid line corresponds to the case
$\gamma_0=0, \ \alpha=0.35$, and dashed line to $\gamma_0=1, \
\alpha=1$ .} \label{gain_q}
\end{figure}

The dependence of MI gain versus the wavenumber of modulations of
plane wave is presented in Fig.~\ref{gain_q}. The dashed line
corresponds to $\gamma =1$,  $\alpha = 1$ and solid line to
$\gamma =0$,  $\alpha = 0.35$. Other parameters of the system are
the same in both cases: $\epsilon = 0.2, \ \kappa = 1, \ f = -1$.


\begin{figure}
\centerline{\includegraphics[width=3.0in,angle=-90]{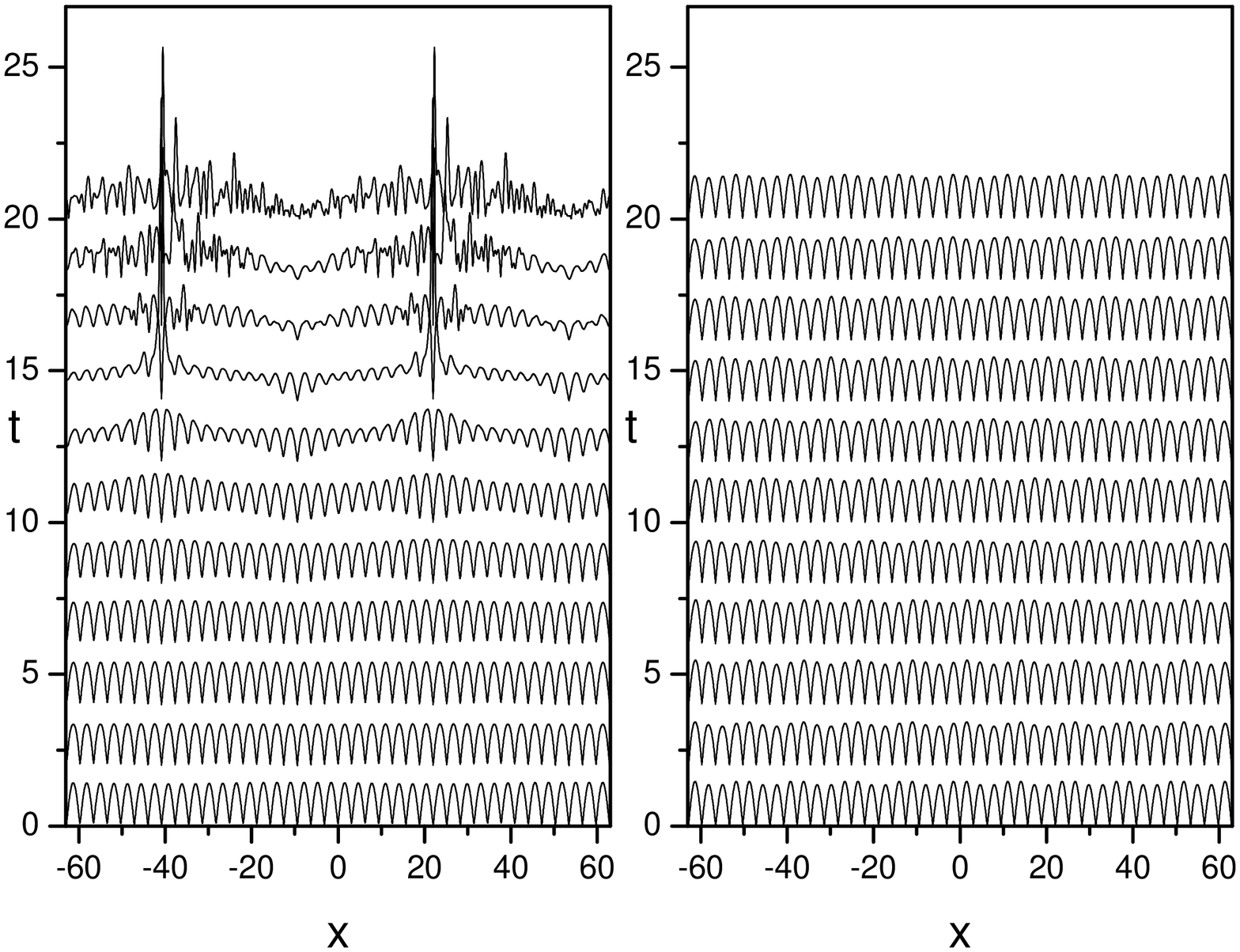}}
\caption{Modulational instability development. Parameters are
$\gamma_0=1, \ \kappa=1, \ \epsilon = 0.2, \ \alpha=1, \ f= -1$.
a) $q=0.1$, b) $q=0.5$} \label{modinst}
\end{figure}

Results of numerical simulations  of MI when $\gamma_0=1,
\kappa=1, \epsilon = 0.2, \alpha=1, f= -1$ is presented in
Fig.~\ref{modinst}. The evolution of amplitudes for  $q=0.1$ and
$q=0.5$ are shown.  Initial value of the wave function amplitude
at $t = 0$ is 1.44. By the time $t = 20$ the amplitude
value grows up to 5.83.\\

\begin{figure}
\centerline{\includegraphics[width=2.5in,angle=-90]{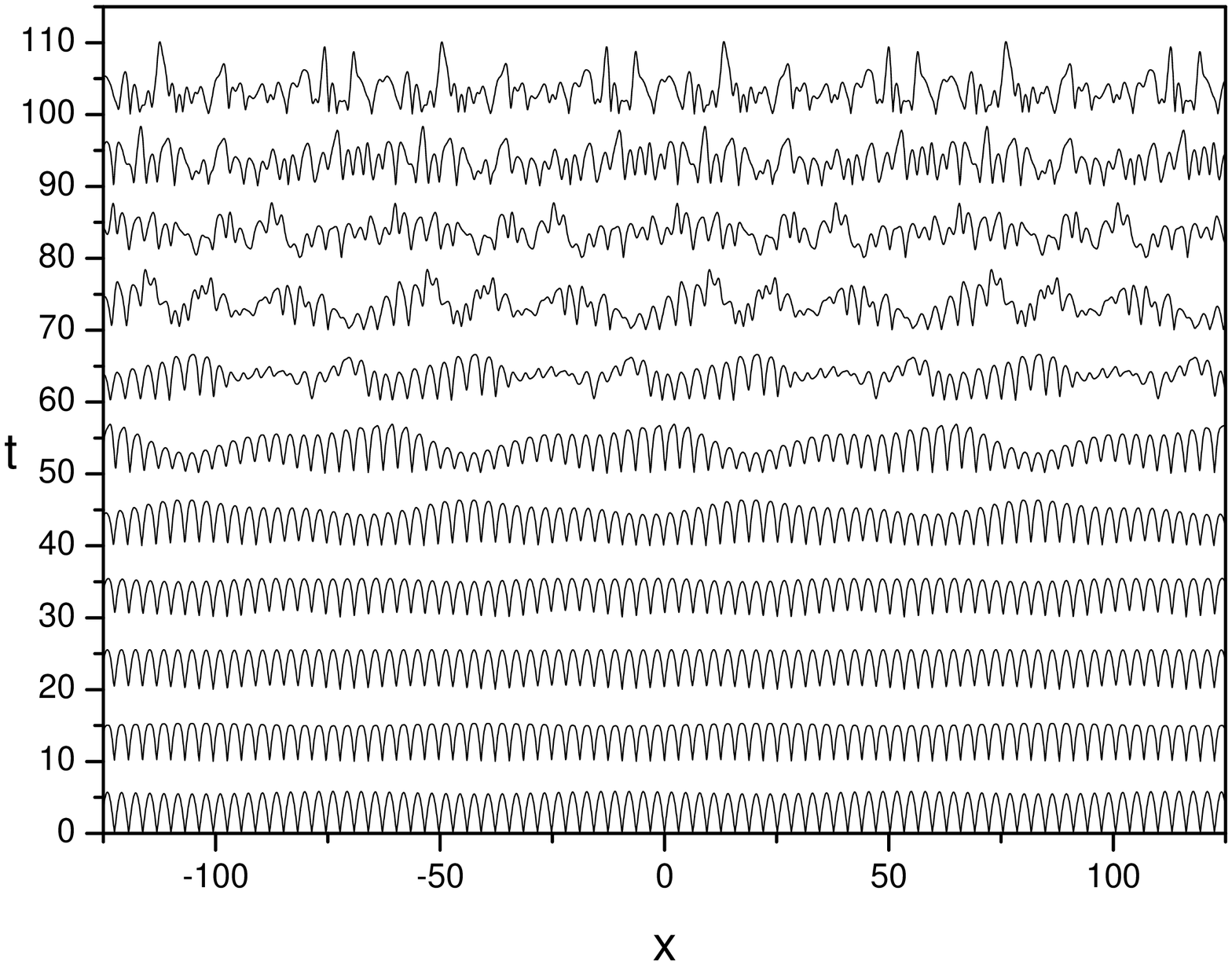}}
\caption{Suppressed modulational instability. Parameters are
$\gamma_0 = 0, \ \kappa=1, \ \epsilon = 0.2, \ \alpha=0.8, \ f= -1
\ q = 0.1$ .} \label{smodinst}
\end{figure}

A curious behaviour of the modulational instability has been
observed when numerical simulating MI at $\gamma = 0$. (Other
parameters of the system were $\kappa=1, \ \epsilon = 0.2$ and
initial cw amplitude $\alpha = 0.8$). The result of simulation is
shown in Fig.~\ref{smodinst}.

\begin{figure}
\centerline{\includegraphics[width=2.5in,angle=-90]{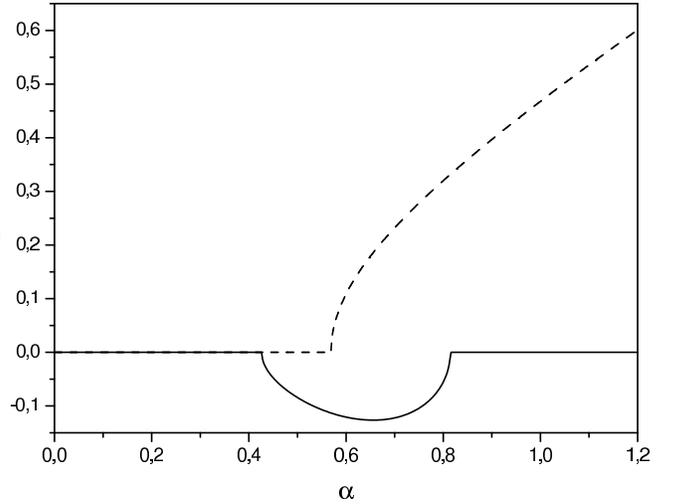}}
\caption{MI gain versus the initial wavefunction amplitude
$\alpha$. Parameters are $\kappa=1, \ \epsilon = 0.2, \
 f= -1$. Solid line corresponds to the case
$\gamma_0=0, \ \alpha=0.8$ and dashed line to $\gamma_0=1, \
\alpha=1$ .} \label{gain-al}
\end{figure}

In this case one can observe suppressed MI evolution with no
growth in the wave function amplitude. Such a development of MI
can be explained from the dependence of MI gain versus the cw
amplitude $\alpha$ given in Fig.~\ref{gain-al}.
As seen, in the case of $\gamma = 0$ MI gain exists in some
limited range $0.42< \alpha < 0.82$. Taking into account that
$\alpha = max|\Psi|$ and that outside of this limited range the
solution is {\it stable}, we can conclude that the growth of the
wave function amplitude must be suppressed.

\begin{figure}
\centerline{\includegraphics[width=2.9in]{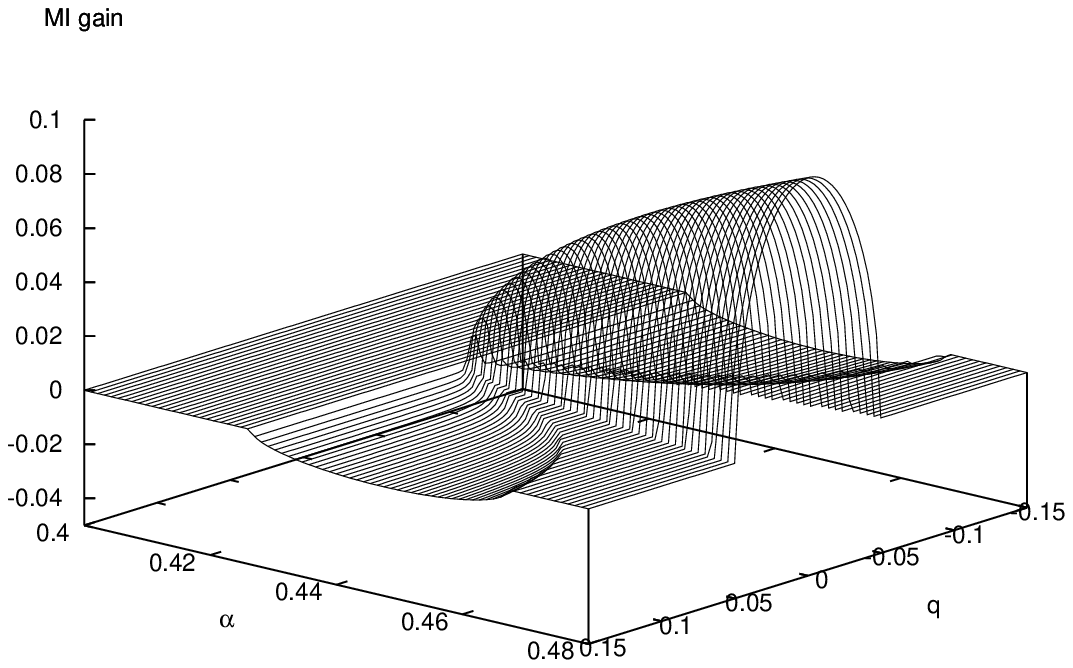}}
\caption{MI gain versus the wavenumber of modulations $q$ and the
initial wavefunction amplitude $\alpha$. Parameters are
$\kappa=1.1, \ \epsilon = 0.2, \  f= -1, \ \gamma_0=1$. (The same
is for $f= 1, \ \gamma_0=-1$.)} \label{gain-1-11}
\end{figure}

\begin{figure}
\centerline{\includegraphics[width=2.9in]{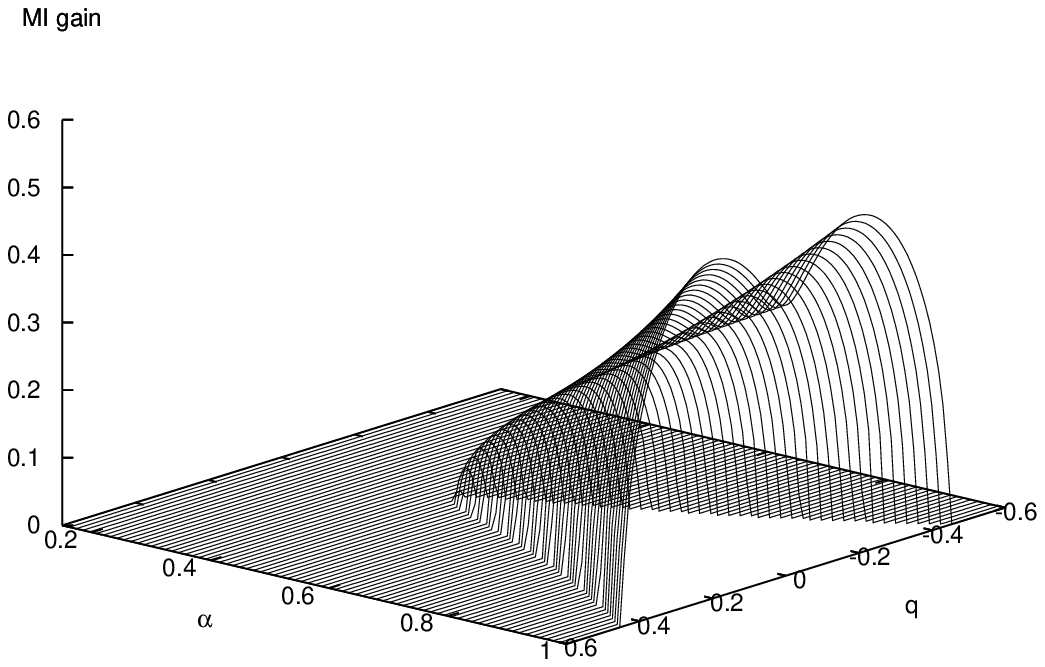}}
\caption{MI gain versus the wavenumber of modulations $q$ and the
initial wavefunction amplitude $\alpha$. Parameters are $\kappa=1,
\ \epsilon = 0.2, \  f= -1, \  \gamma_0=1$. (The same is for $f=
1, \ \gamma_0=-1$.)} \label{gain-1-1}
\end{figure}

In general dependence of MI gain on the system parameters is very
complicated that is demonstrated in Figs.~\ref{gain-1-11} -
\ref{gain-1-1}.  Results of numerical simulations confirm
predictions of the theory.
\begin{figure}
\centerline{\includegraphics[width=2.9in]{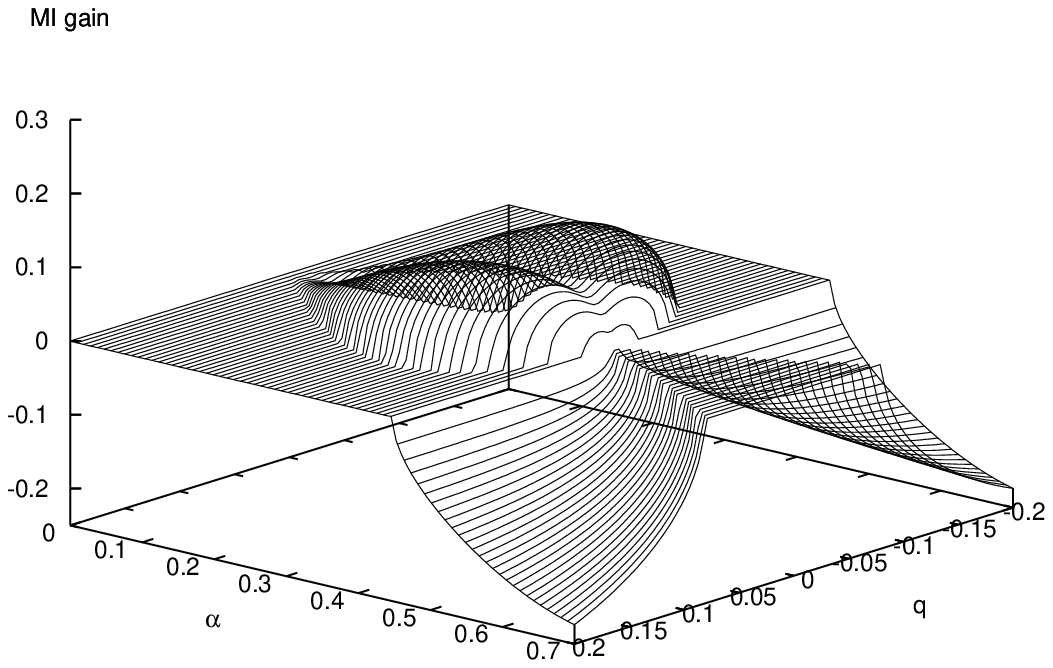}}
\caption{MI gain versus the wavenumber of modulations $q$ and the
initial wavefunction amplitude $\alpha$. Parameters are $\kappa=1,
\ \epsilon = 0.2, \ f= -1, \ \gamma_0=-1$. (The same is for $f= 1,
\ \gamma_0=1$.) } \label{gain-m1-m1}
\end{figure}


\section{Gap soliton in linear-nonlinear lattice}
 To find gap soliton solution let us look for the particular
solution of the form $A,B \rightarrow (a(x),b(x))\exp(-i\omega
t)$.



The symmetry of the system (\ref{cm}) allows to consider the case
$a(x)=b^{\ast}(x)$\cite{pelinovsky}. Then we get the equation for
$a(x)$
  \begin{eqnarray}\label{eq5}
   \omega a+2ia_x+3\gamma_0|a|^2 a-\frac{\epsilon}{2}a^{\ast}+
   \nonumber\\
   \frac{3}{2}\kappa |a|^2 a^{\ast}+\frac{\kappa}{2}a^3 = 0.
  \end{eqnarray}
The solution will be sought in the form
$$a(x) = \sqrt{U(x)}e^{-i\theta(x)/2}.$$
Substituting this expression into Eq.(\ref{eq5}) we get the
following system of equations
  \begin{eqnarray}\label{eq19}
   \theta_x = -\omega -3\gamma_0 U + \frac{\epsilon}{2}\cos{\theta} -
   2\kappa U \cos{\theta},\\\label{eq20}
   U_x = \frac{\epsilon}{2}U\sin{\theta} -\kappa U^2 \sin{\theta}.
  \end{eqnarray}
The system has the Hamiltonian $H_g$ and can be written in the
form
  \begin{equation}
   \theta_x = \frac{\partial H_g}{\partial U}, \quad
   U_x = -\frac{\partial H_g}{\partial\theta},
  \end{equation}
where
  \begin{equation}
   H_g = -\omega U - \frac{3}{2}\gamma_0 U^2 +
   \frac{\epsilon}{2}U\cos{\theta}-\kappa U^2\cos{\theta}.
  \end{equation}

\begin{figure}
\centerline{\includegraphics[width=3in]{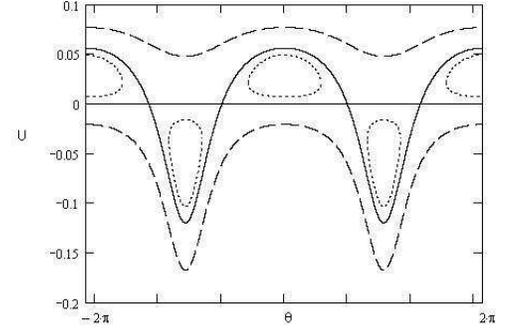}}
\caption{Phase-plane portrait $U=U(\theta,H_g )$ at $\gamma_0 =
1$, $\epsilon = 0.2$, $\kappa = 1$ and $\omega = -0.04$. Solid
line is for $H = 0$, dashed line for $H = -0.04$, dotted line for
$H = 0.00085$. } \label{phase}
\end{figure}

We will consider solutions that rapidly decay in the space, i.e.
$U \rightarrow 0, |x| \rightarrow \infty$. In this case $H_g = 0$,
that corresponds to separatrix in the phase-plane portrait (see
Fig. \ref{phase}).

Then  for $\theta$ we have the following equation
  \begin{equation}\label{eq9}
   \theta_x = \omega - \frac{\epsilon}{2}\cos{\theta},
  \end{equation}
with $U$ determined as
  \begin{equation}\label{eq10}
   U=-\frac{2\omega-\epsilon\cos{\theta}}{3\gamma_0+2\kappa\cos{\theta}}.
  \end{equation}
Integrating Eq.(\ref{eq9}) provided that $|\tan(\theta/2)|<\gamma$
we find
  \begin{equation}
   \cos{\theta} = \frac{1 - \gamma^2 \tanh^2(\beta x)}{1 +
   \gamma^2\tanh^2(\beta x)},
  \end{equation}
where
  $$
    \gamma = \sqrt{\frac{\epsilon - 2\omega}{\epsilon + 2\omega}},
    \quad
    \beta  = \frac{\sqrt{\epsilon^2 - 4\omega^2}}{4}.
  $$

From Eq.(\ref{eq10}) we obtain expression for the amplitude U
  \begin{equation} \label{U}
   U(x)=\frac{\epsilon -2\omega}
   {(3\gamma_0+2\kappa)\cosh^2(\beta x)+
   (3\gamma_0 -2\kappa) \gamma^2 \sinh^2(\beta x)}.
  \end{equation}

Finally the gap solution takes the form
  \begin{equation}\label{sol1}
   u(x,t) = 2\sqrt{U}
   \cos\left(x-\frac{\theta(x)}{2}\right)e^{-i(\omega +1 )t},
  \end{equation}
with
  \begin{equation} \label{omega}
   -\epsilon/2 < \omega < \epsilon/2 .
  \end{equation}

The number of atoms in the soliton is
  \begin{equation}
   N=\int_{-\infty}^{\infty}|u(x,t)|^2 dx=2\int_{-\infty}^{\infty}U dx.
  \end{equation}
When $3\gamma_0> 2\kappa$
  \begin{equation}
   N=\frac{8}{\sqrt{9\gamma_0^2-4\kappa^2}}\arctan
   \frac{(3\gamma_0-2\kappa)\gamma}{\sqrt{9\gamma_0^2-4\kappa^2}},
  \end{equation}
and when
$3\gamma_0< 2\kappa$
  \begin{equation}
   N=\frac{8}{\sqrt{4\kappa^2-9\gamma_0^2}}\ln\left|
   \frac{3\gamma_0+2\kappa+\gamma \sqrt{4\kappa^2-9\gamma_0^2}}
   {3\gamma_0+2\kappa-\gamma \sqrt{4\kappa^2-9\gamma_0^2}}\right|
  \end{equation}

In the case
  \begin{equation} \label{gap}
3\gamma_0 > 2 \kappa
  \end{equation}
solution (\ref{U}) is stable for all values of $\omega$ within
condition (\ref{omega}), i.e. the soliton exists everywhere in the
gap.

In the case $3\gamma_0 < 2\kappa$ for stable solutions we have
another condition for $\omega$
  \begin{equation} \label{omega0}
   -\frac{3\epsilon\gamma_0}{4\kappa}<\omega<\frac{\epsilon}{2}.
  \end{equation}

Chemical potential $\omega$ versus norm $N$ is depicted in
Fig.{\ref{norm}}.
\begin{figure}
\centerline{\includegraphics[width=2.0in,angle=-90]{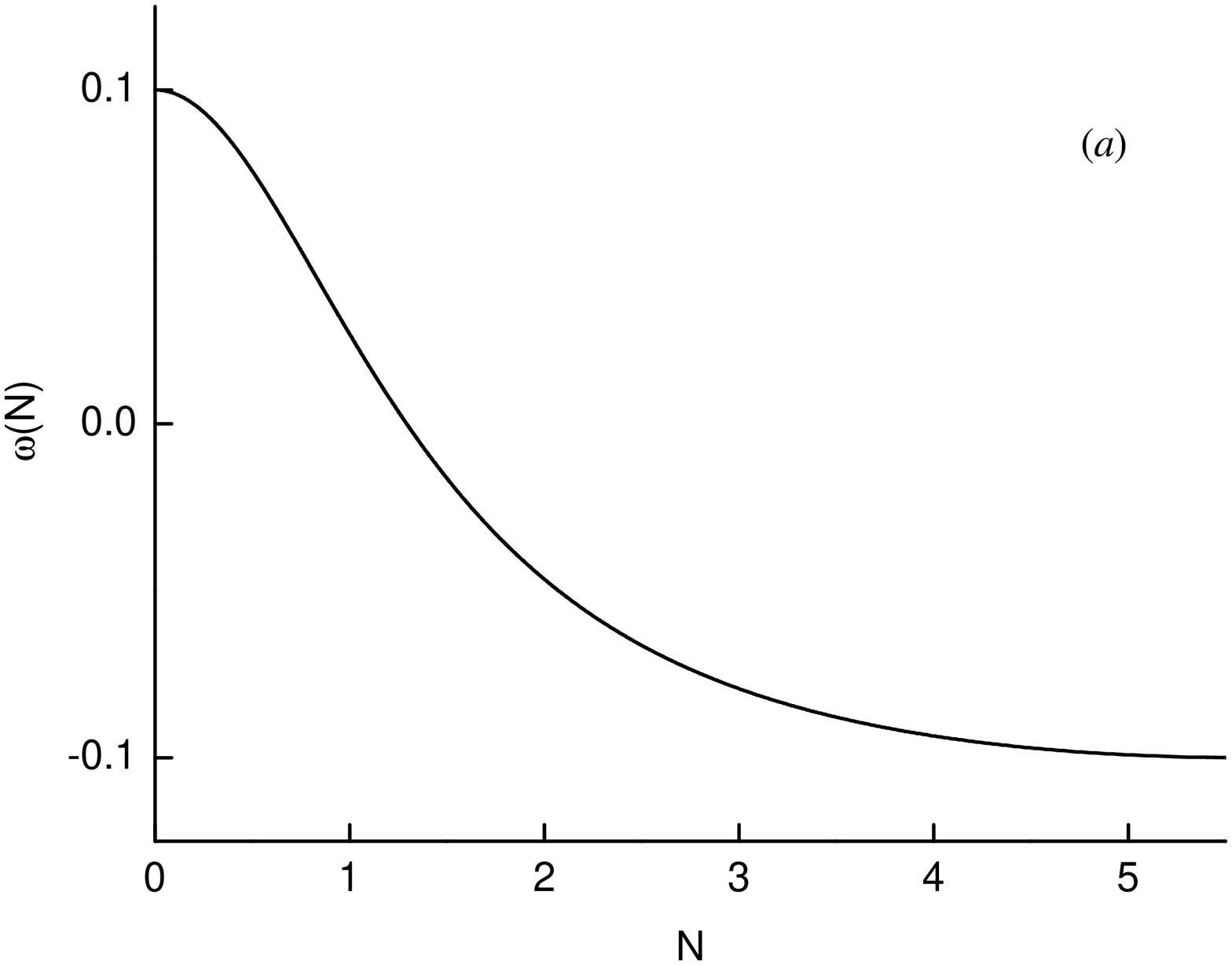}}
\centerline{\includegraphics[width=2.0in,angle=-90]{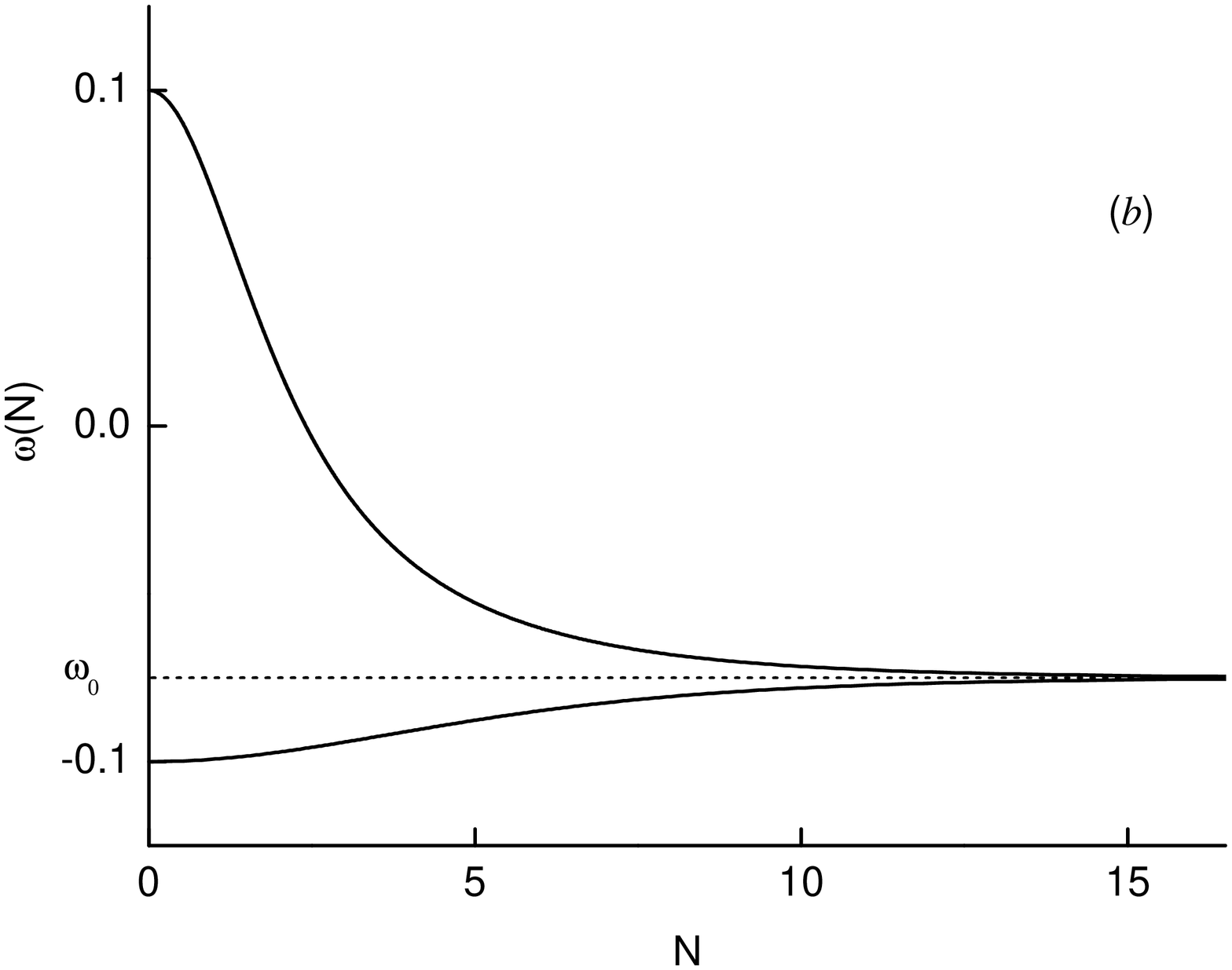}}
\caption{The chemical potential $\omega$ versus norm $N$. The
curve (a) corresponds to condition (\ref{gap}) and the curve (b)
to (\ref{omega0}).} \label{norm}
\end{figure}
According to the Vakhitov-Kolokolov criterion\cite{VK} the soliton
is stable provided that $dN/d\omega < 0$.

\begin{figure}
   \centerline{\includegraphics[width=2.4in,angle=-90]{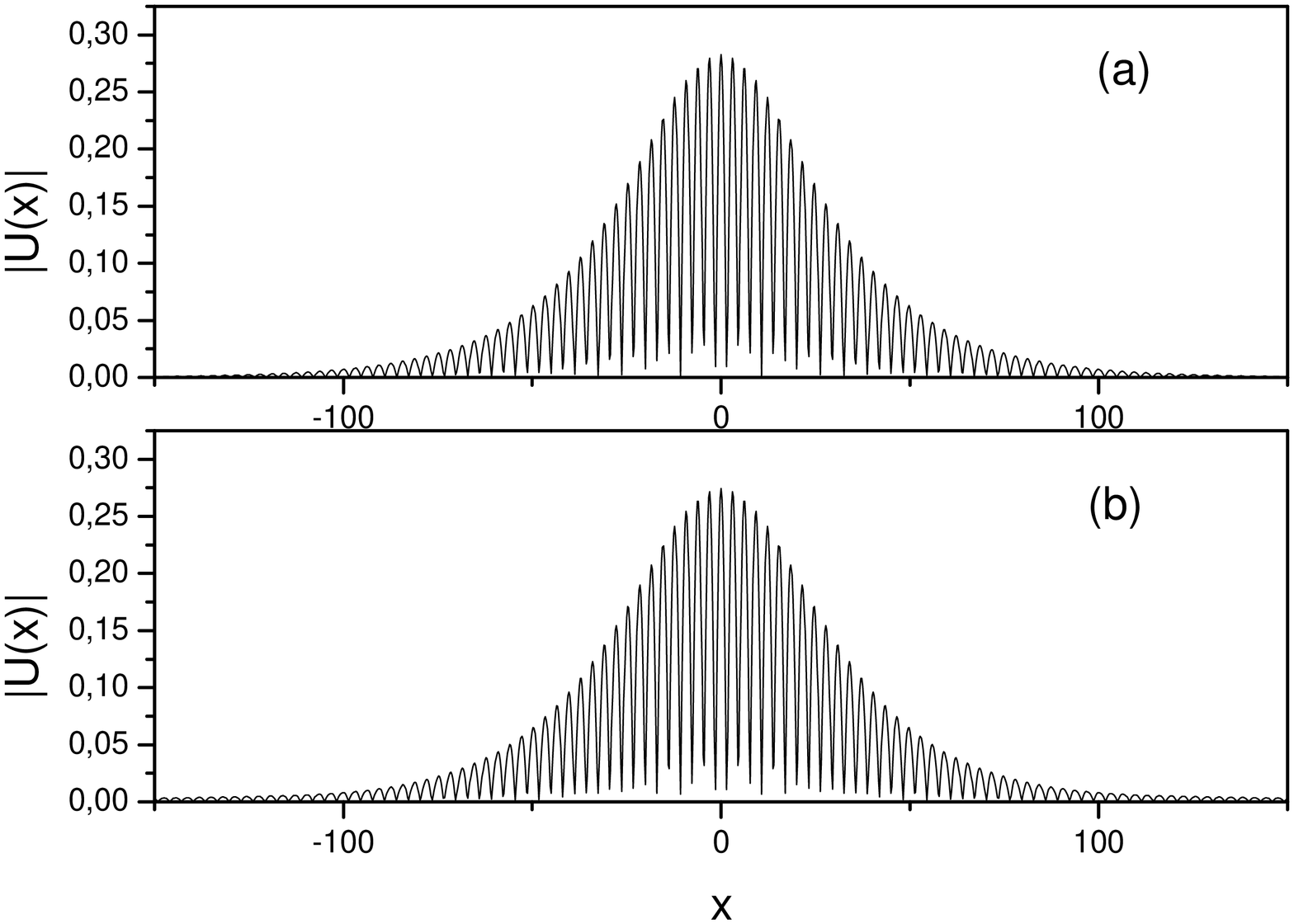}}
   \caption{Stable solution for $\gamma_0 = 1$, $\epsilon = 0.2$,
   $\kappa = 1$ and $\omega = 0.05$ at different times : (a) $t=0$
   and (b) $t=100$. The initial wave packet envelope is taken in the
   form of Eq.(\ref{U}).} \label{stg1}
  \end{figure}

Evolution of the stable solution for $\gamma_0 = 1$, $\epsilon =
0.2$ is given in Fig. \ref{stg1}, The initial wave packet envelope
is taken in the form of Eq.(\ref{U}).

\begin{figure}
   \centerline{\includegraphics[width=2.4in,angle=-90]{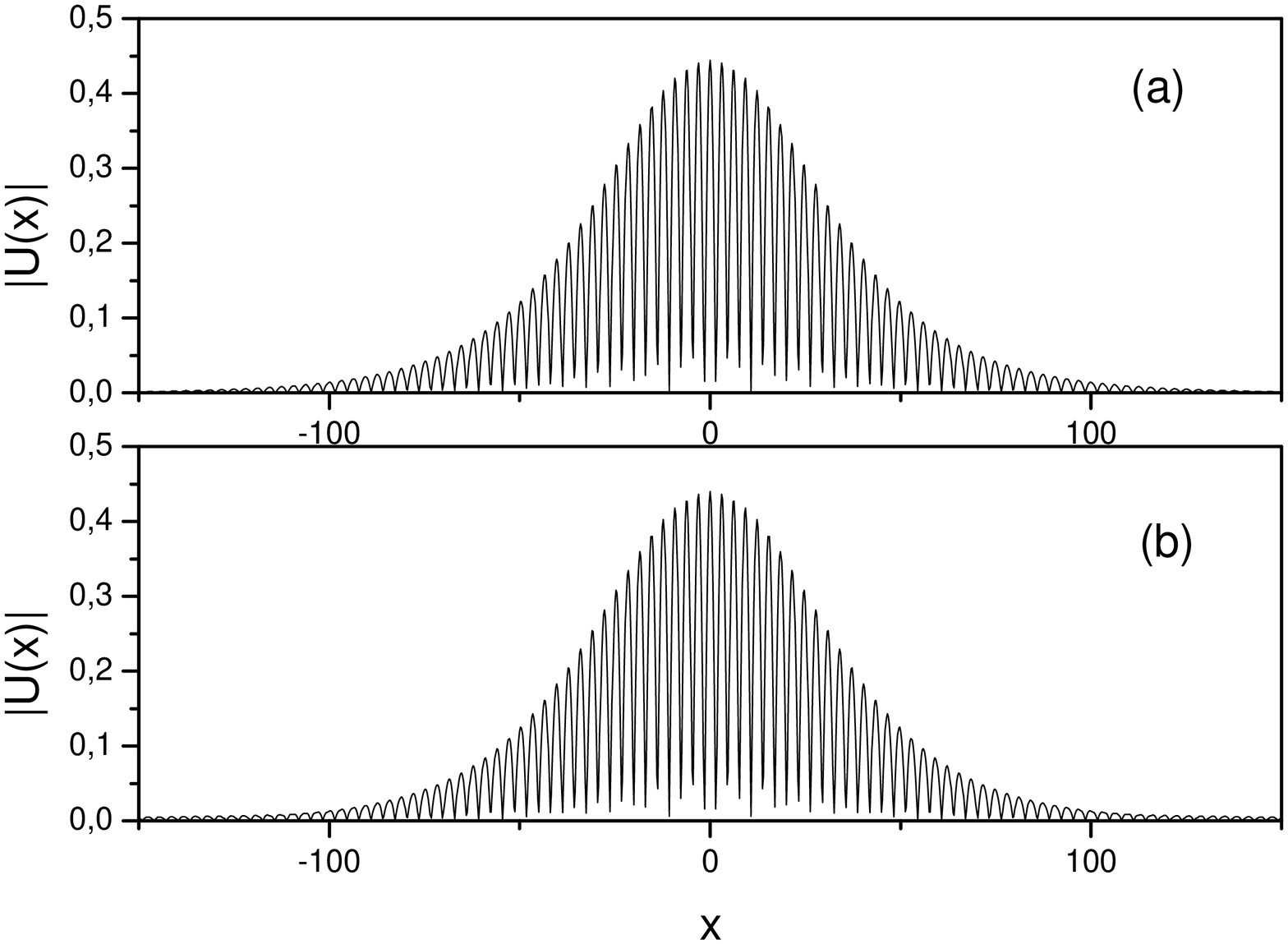}}
   \caption{Stable solution for $\gamma_0 = 0$, $\epsilon = 0.2$,
   $\kappa = 1$ and $\omega = 0.05$ at different times : (a) $t=0$
   and (b) $t=100$. The initial wave packet envelope is taken in the
   form Eq.(\ref{U}).} \label{stg0}
  \end{figure}

An interesting case we have when $\gamma_0 = 0$. It can be
realized with the Feshbach resonance method. Then the soliton
exists in the interval $0 < \omega < \epsilon/2$, i.e. in the
upper half of the gap.  The number of atoms for $\omega = \Delta
\ll \epsilon $ near the center of the gap can be estimated as
$$N
\approx \frac{4}{\pi}\ln|\frac{\epsilon}{\Delta}|.$$ When $\omega
= \epsilon/2 - \Delta$, i.e. the parameter near the band edge, the
number of atoms in gap soliton tends to zero like $$N \rightarrow
\frac{4}{\pi}\sqrt{\frac{\Delta}{\epsilon}}.
$$ Numerical simulation of this case is shown in Fig.
\ref{stg0}.

We performed numerical simulations of the GP equation with the
linear-nonlinear optical potentials. In Figs. \ref{stg1},
\ref{stg0}  evolution of the initial state in Eq.(\ref{GP}) is
given in the form of Eq.(\ref{U}).

\begin{figure}
\centerline{\includegraphics[width=2.4in,angle=-90]{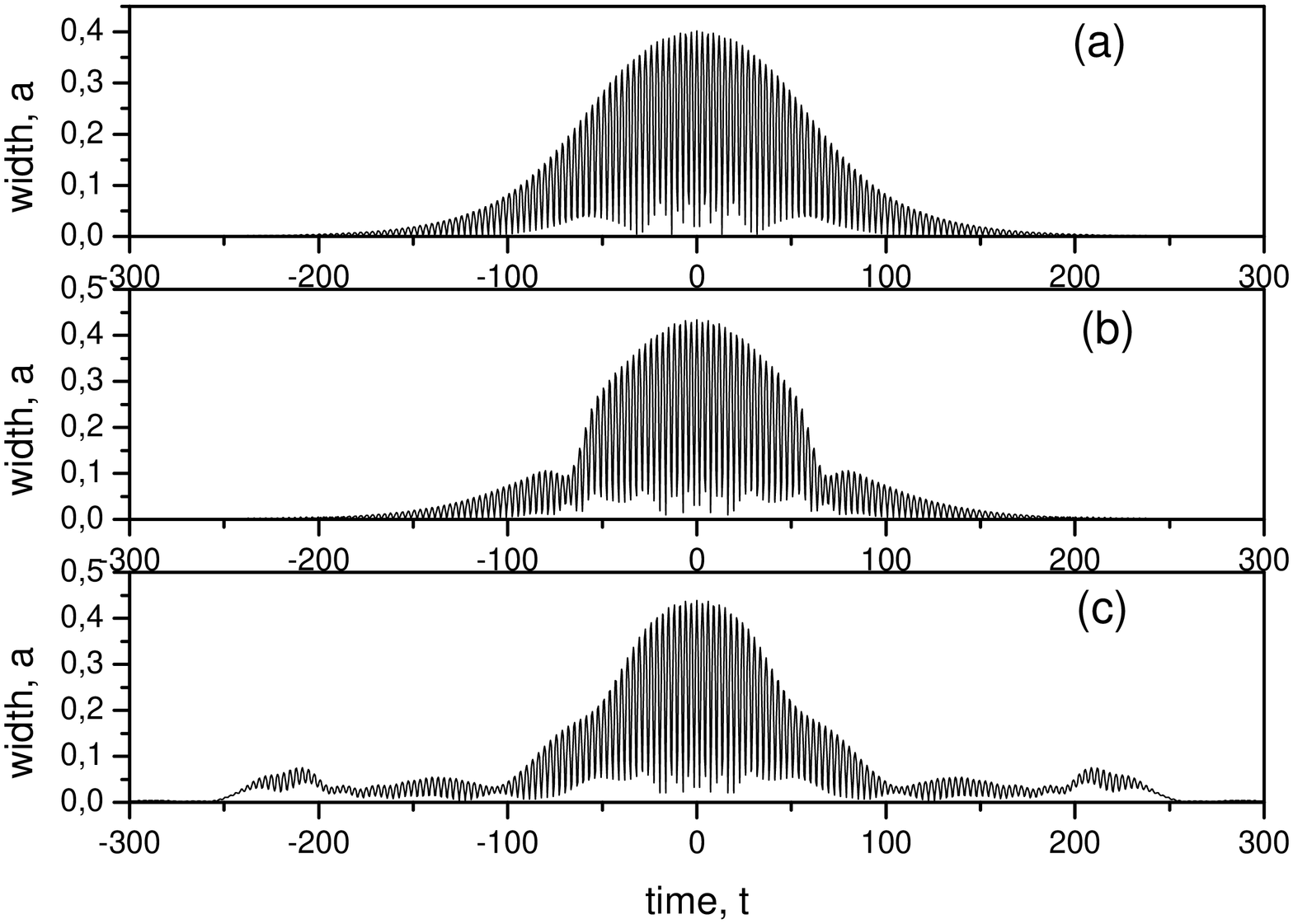}}
\caption{Unstable solution for $\gamma_0 = 1$, $\epsilon = 0.2$,
$\kappa = 2$ and $\omega = -0.08$ at different times : (a) $t=0$,
(b) $t=40$ and (c) $t=140$. The initial wave packet envelope is
taken in the form Eq.(\ref{Us}).} \label{unst}
\end{figure}

When $|\tan(\theta/2)|> \gamma$ the expressions for the amplitude
$U$ and $\cos\theta$ take the forms
  \begin{equation}
   \cos{\theta} = \frac{   \tanh^2(\beta x)-\gamma^2}
   {\tanh^2(\beta x)+\gamma^2},
  \end{equation}
  \begin{equation} \label{Us}
   U(x) = -\frac{\epsilon-\omega}
   {(3\gamma_0+2\kappa)\sinh^2(\beta x)+
   (3\gamma_0 -2\kappa) \gamma^2\cosh^2(\beta x)}.
  \end{equation}
In this case the solution is evidently unstable. Indeed, in this
case $dN/d\omega > 0$ and in accordance with the
Vakhitov-Kolokolov criterion\cite{VK} such a solution is unstable.
PDE simulation of the evolution of an unstable solution for
$\gamma_0 = 1$, $\epsilon = 0.2$ and $\kappa = 2$ is given in Fig.
\ref{unst}, The initial wave packet envelope is taken in the form
of Eq.(\ref{Us}).

\section{Travelling gap-soliton solution. Solitons collision}
Travelling solitons can be obtained proceeding from the anzatz of
work\cite{Iizuka}. The  solution of Eqs.(\ref{cm}-\ref{cm1})
should be sought in the form of the sum of travelling waves with
different amplitudes and phases
  \begin{eqnarray} \label{wave_int}
   A(x,t)=\sqrt{\Delta}\sqrt{F_A(\xi)}e^{i(\theta_A(\xi)-\omega
   t)},   \\
   B(x,t)=\frac{1}{\sqrt{\Delta}}\sqrt{F_B(\xi)}e^{i(\theta_B(\xi)-\omega
   t)},
  \end{eqnarray}
  where
    $$
        \Delta=\sqrt{\frac{2+v}{2-v}}, \ \xi = x - vt.
    $$
Substituting these expressions into Eqs.(\ref{cm}-\ref{cm1}) we
have
  \begin{eqnarray} \label{wave}
   F_\xi = \frac{\widetilde{\epsilon}}{2} F\sin({\theta_B-\theta_A}) -
   \widetilde{\kappa} F^2 \sin({\theta_B-\theta_A}),\\ \label{wave1}
   (\theta_B-\theta_A)_\xi = -\widetilde{\omega} +
   \frac{\widetilde{\epsilon}}{2}\cos({\theta_B-\theta_A})-
   \nonumber \\
   3\widetilde{\gamma_0}F -
   2\widetilde{\kappa} F \cos({\theta_B-\theta_A}),\\ \label{wave2}
   (\theta_B+\theta_A)_\xi =\frac{v}{2}\Bigl[\widetilde{\omega}+
   \frac{8}{4+v^2}\widetilde{\gamma}_0 F+\nonumber \\
   \widetilde{\kappa} F\cos(\theta_B-\theta_A)\Bigr],
  \end{eqnarray}
where $F=F_A=F_B$ ,
  \begin{eqnarray} \label{tilde}
   \widetilde{\omega}=\lambda^2 \omega,\quad
   \widetilde{\epsilon}=\lambda\epsilon,\quad
   \widetilde{\kappa}=\lambda^2 \kappa, \nonumber \\
   \widetilde{\gamma}_0=\frac{(4+v^2)}{4}\lambda^3\gamma_0, \quad
   \lambda=\frac{1}{\sqrt{1-\ds\frac{v^2}{4}}}.
  \end{eqnarray}

The first two equations of this system coincide with Eqs.
(\ref{eq19}-\ref{eq20}) with renormalized parameters. Thus the
region of the solitons stability are determined again by
conditions Eqs.(\ref{omega}), (\ref{omega0}) with renormalized
parameters as in Eq.(\ref{tilde}).

The gap is given by
\begin{equation}
-\frac{\epsilon }{2}\sqrt{1 - \frac{v^2}{4}} < \omega <
\frac{\epsilon}{2} \sqrt{1 - \frac{v^2}{4}}.
\end{equation}
When the velocity tends to the limiting value $v \rightarrow 2$,
the gap region shrinks to zero. Also the gap does not exist for
the velocity $v > 2$.

The stability condition is
\begin{equation}
 -\frac{3(4+v^2)\epsilon\gamma_0}{4\kappa}   < \omega <
 \frac{\epsilon}{2}\sqrt{1-\frac{v^2}{4}}.
\end{equation}

 Numerical simulations with the
initial wave packet taken in the form
  \begin{eqnarray} \label{movesol}
   u(x,t)=(A(\xi)e^{ix}+B(\xi)e^{ix})e^{-i(\omega+1)t}
  \end{eqnarray}
show that obtained solution of Eq.(\ref{GP}) has the form of an
automodel soliton.

\begin{figure}
\centerline{\includegraphics[width=2.4in,angle=-90]{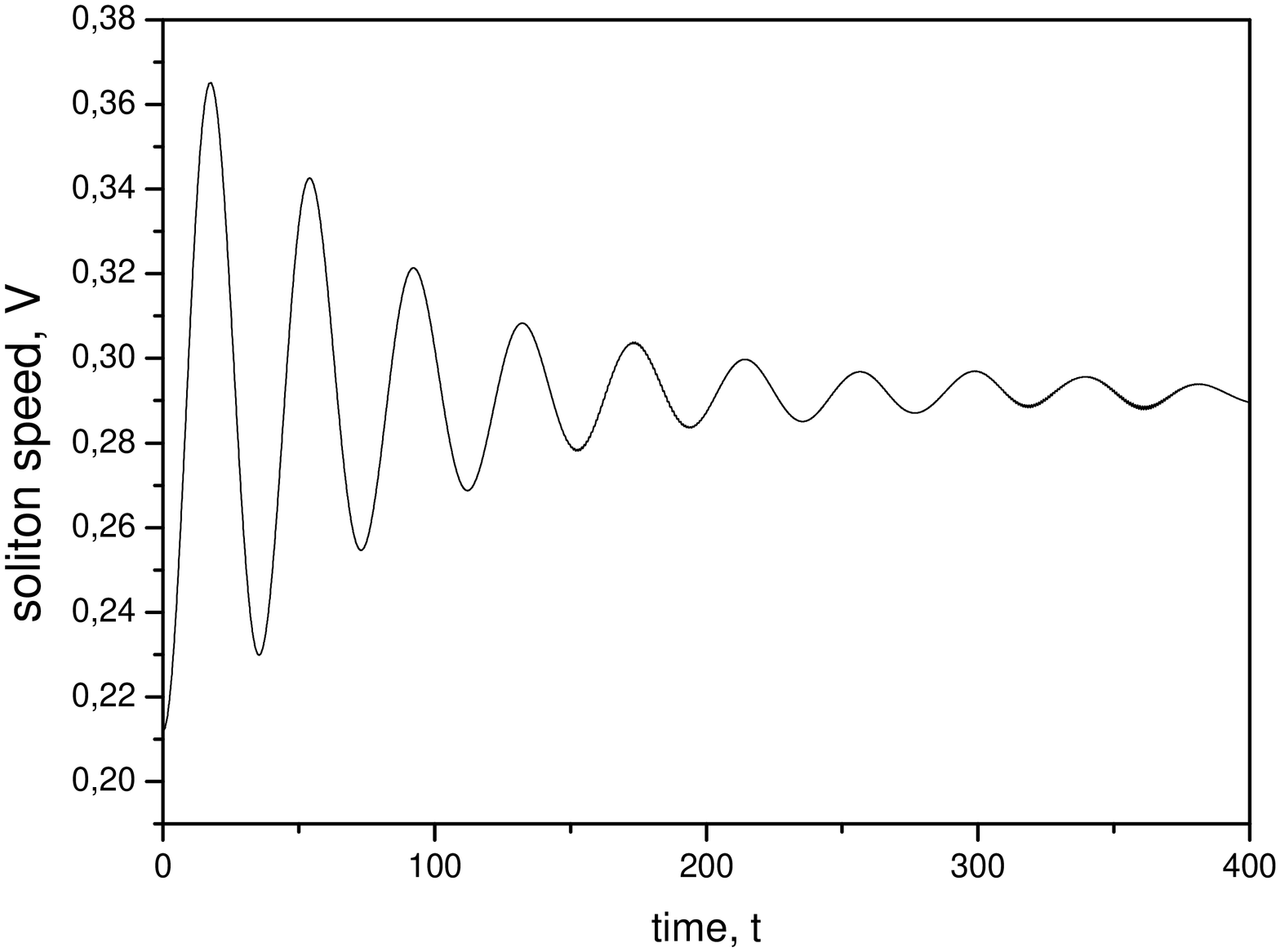}}
\caption{The time dependence of the travelling soliton speed for
$\gamma_0 = 1$, $\epsilon = 0.2$, $\kappa = 1$ and $\omega = 0.05$
when the initial speed $v = 0.2$.} \label{speed}
\end{figure}

In the course of evolution of the initial wave-packet the soliton
parameters, the soliton speed, amplitude and width, approach
their stationary values. Evolution of the soliton speed is shown
in Fig. \ref{speed}.

\begin{figure}
\centerline{\includegraphics[width=2.4in,angle=-90]{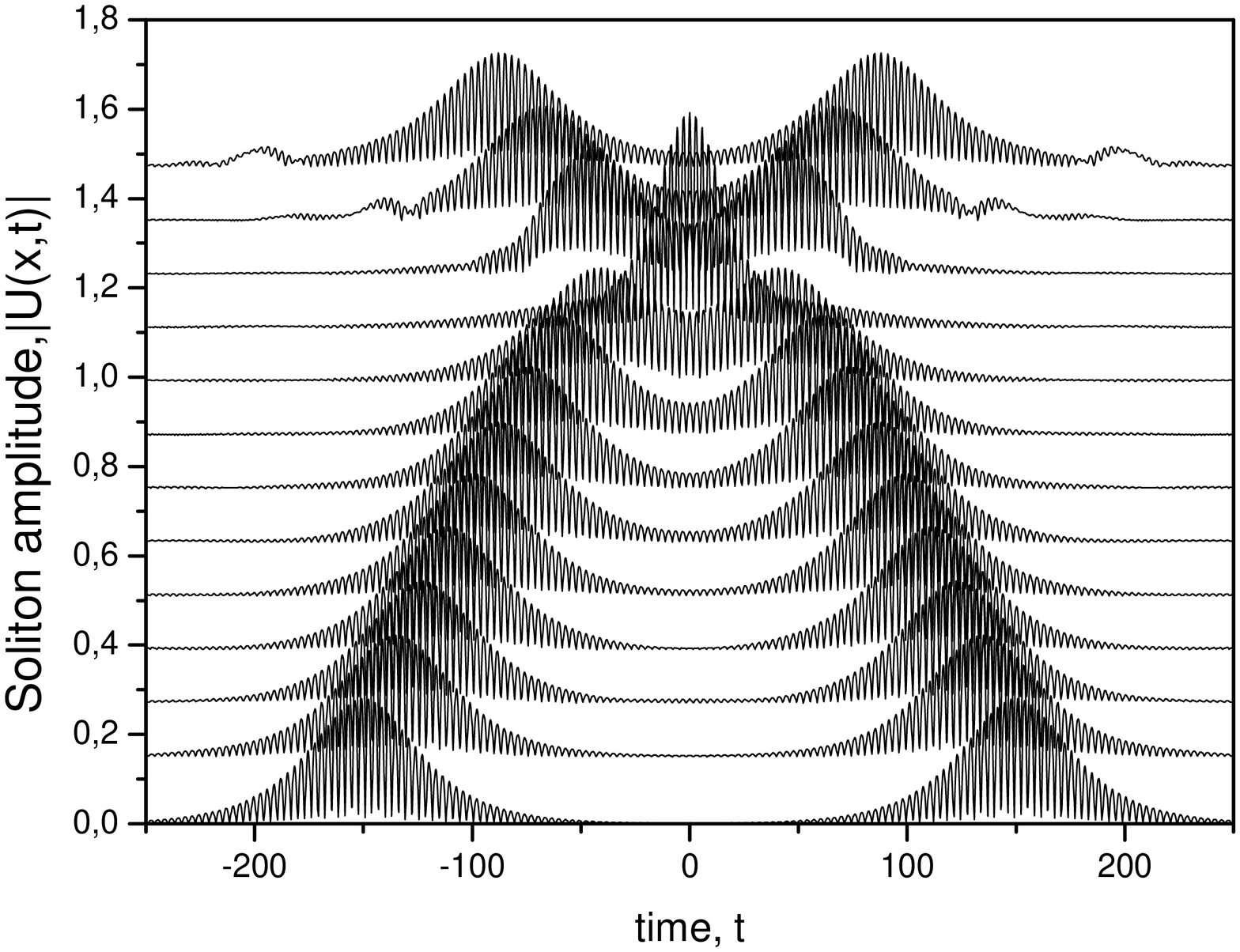}}
\caption{Collision of two moving solitons for $\gamma_0 = 1$,
$\epsilon = 0.2$, $\kappa = 1$ and $\omega = 0.05$.}
\label{collis}
\end{figure}

%
To check the nature of obtained moving solutions in Fig.
\ref{collis} the result of simulation of two solitary waves
collision taken in the form of Eq.(\ref{movesol}) is shown.
As is evident from this figure, collision of two solitons is
nearly elastic. A small radiation at the solitons tails seems to
be caused by the nonintegrability of the system.

\section{Conclusion}
In conclusion, we have investigated the dynamics of BEC under
joint action of the linear and nonlinear optical lattices. These
lattices can be created by counter propagating laser beams via
optically induced Feshbach resonances. Also such structures can be
generated in arrays of the Fermi-Bose mixtures. For the case of
shallow lattices we derive the coupled mode system of equations
and have investigated the modulational instability of cw matter
waves in such periodic media. We  find the instability regions of
cw nonlinear matter waves. The knowledge of MI regions can be
useful for generation of gap solitons in such structures. New
types of gap soliton solutions are found and the stability regions
are analyzed. In particular, the gap soliton can exist in the
upper half of the forbidden gap,  when the background atomic
scattering length is equal to zero($\gamma_0 = 0$). We find the
travelling gap soliton solution in BEC and study collision of gap
solitons by numerical simulations. It is shown that the collisions
are nearly elastic, with the emitting of small radiation, due to
the nonintegrability of the couple-mode system.


\begin{acknowledgments}
The work was partially supported by the Fund for Fundamental
Research support of the Uzbek Academy of Sciences (Award 20-06).
F.Kh.A. is grateful to FAPESP (Sao Paulo, Brasil) for the partial
support. Authors are grateful to B.B. Baizakov and M. Salerno for
useful discussions.
\end{acknowledgments}


\newpage
\begin{thebibliography}{99}
\bibitem{Morsh}
O. Morsh and M. Oberthaler, Rev.Mod.Phys. {\bf 78}, 179 (2006).

\bibitem{BK}
V.A. Brazhnyi and V.V. Konotop, Mod.Phys.Lett. B {\bf 18}, 627
(2004).

\bibitem{AS03}
F.Kh. Abdullaev and M. Salerno, J.Phys. B {\bf 36}, 2851 (2003).

\bibitem{kevrekidis}
G. Theocharis, P. Schmelcher, P.G. Kevrekidis, and D.J.
Frantzeskakis, Phys.Rev. A {\bf 72}, 033614 (2005).

\bibitem{AGKT}
F.Kh. Abdullaev, A. Gammal, A.M. Kamchatnov, and L. Tomio, Int.J.
Mod.Phys. B {\bf 19}, 3415 (2005).


\bibitem{GA}
J. Garnier and F.Kh. Abdullaev, Phys.Rev. A {\bf 74}, 013604
(2006).


\bibitem{SM}
H. Sakaguchi and B.A. Malomed, Phys.Rev. E {\bf 72}, 046610
(2005).

\bibitem{AG}
F.Kh. Abdullaev and J. Garnier, Phys.Rev. A {\bf 72} 061605(R)
(2005).


\bibitem{Fibich}
G. Fibich, Y. Sivan, and M.I. Weinstein, Physica D {\bf 217}, 31
(2006).

\bibitem{OFR1}
M. Theis, G. Thalhammaer, K. Winkler, M. Hellwig, G. Ruff, R.
Grimm, and J.H. Denshlag, Phys.Rev.Lett. {\bf 93}, 123001 (2004).


\bibitem{OFR2}
P.O. Fedichev, Yu. Kagan, G.V. Shlyapnikov, and J.T.M. Walraven,
Phys.Rev.Lett. {\bf 77}, 2913 (1996).

\bibitem{Konotop06}
Yu. Bludov and V.V. Konotop, arXive cond-mat/0606815.

\bibitem{Malomed05}
B.A. Malomed, T. Mayteevarunyoo, E. Ostrovskaya, and Y.S. Kivshar,
Phys.Rev.E {\bf 71}, 056616 (2005).

\bibitem{AW}
A.B. Aceves and S. Wabnitz, Phys.Lett. A {\bf 141}, 37 (1989);
D.N. Christodoulides and R.I. Joseph, Phys.Rev.Lett. {\bf 62},
1746 (1989).

\bibitem{Iizuka}
T. Iizuka and C.M. de Sterke, Phys.Rev. E {\bf 61}, 4491 (2000).

\bibitem{Pel1} D. Pelinovsky, J. Sears, L. Brzozowski, and E.H.
Sargent, J.Opt.Soc.Am.  B {\bf 19}, 43 (2002).

\bibitem{Sterke}
C.M. de Sterke and J.E. Sipe,  Progress in Optics, {\bf 33}, 203
(1994).

\bibitem{YS}
A.V. Yulin and D.V. Skryabin, Phys.Rev. A {\bf 67}, 023611 (2003).

\bibitem{EC}
N.K. Efremidis and C.N. Christodoulides, Phys.Rev. A {\bf 67},
063608  (2003).

\bibitem{Eggleton}
B.J. Eggleton, R.R. Slusher, C.M. de Sterke, P.A. Krug, and J.E.
Sipe, Phys.Rev.Lett. {\bf 76}, 1627 (1996).


\bibitem{TS}
A. Trombettoni and A. Smerzi, Phys.Rev.Lett. {\bf 86}, 2353
(2001).

\bibitem{ABDKS}
F.Kh. Abdullaev, B.B.Baizakov, S.A. Darmanyan, V.V. Konotop, and
M. Salerno, Phys.Rev. A {\bf 64}, 043606 (2001).


\bibitem{Khare}
A. Khare, K.O. Rasmussen, M. Salerno, M.R. Samuelsen, and A.
Saxena, Phys.Rev.E {\bf 74}, 016607 (2006).

\bibitem{ADG}
F.Kh. Abdullaev, S.A. Darmanyan, and J. Garnier, In Progress in
Optics, {\bf 44}, 303 (2002).

\bibitem{KS}
V.V. Konotop and M. Salerno, Phys.Rev.A {\bf 65}, 021602 (2002).


\bibitem{Sterke2}
C.M. de Sterke, J.Opt.Soc.Am. B 15, 2660 (1998).

 \bibitem{Porz} K. Porsezian and K. Senthilnathan, Chaos  {\bf
15}, 037109 (2005).

\bibitem{pelinovsky}
M. Chugunova and D. Pelinovsky, SIAM J. Appl. Dyn. Sys. {\bf 5},
66 (2006).

\bibitem{VK}
N.G. Vakhitov and A.A. Kolokolov, Radiophysics and Quantum
Electronics, {\bf 16}, 783 (1973).

\bibitem{AS05}
F.Kh. Abdullaev and M. Salerno, Phys.Rev. A {\bf 72}, 033617
(2005).







\end{thebibliography}
\end{document}